\documentclass[reprint,prl,showpacs,superscriptaddress,english]{revtex4-1}
\usepackage[utf8]{inputenc}
\usepackage{color}
\usepackage{array}
\usepackage{multirow}
\usepackage{amsmath}
\usepackage{amssymb}
\usepackage{graphicx}
\usepackage{esint}
\usepackage{pdfpages}
\usepackage[unicode=true,
 bookmarks=false,
 breaklinks=true,pdfborder={0 0 0},backref=false,colorlinks=true]
 {hyperref}

\makeatletter

\@ifundefined{textcolor}{}
{%
 \definecolor{BLACK}{gray}{0}
 \definecolor{WHITE}{gray}{1}
 \definecolor{RED}{rgb}{1,0,0}
 \definecolor{GREEN}{rgb}{0,1,0}
 \definecolor{BLUE}{rgb}{0,0,1}
 \definecolor{CYAN}{cmyk}{1,0,0,0}
 \definecolor{MAGENTA}{cmyk}{0,1,0,0}
 \definecolor{YELLOW}{cmyk}{0,0,1,0}
}

\usepackage{graphicx}
\usepackage{epstopdf}\usepackage{subfigure}
\usepackage{float}
\usepackage{amsfonts}
\usepackage{bm}
\usepackage{subfigure}
\usepackage{caption}

\makeatother

\begin{document}

\title{Detecting Chiral Magnetic Effect by Lattice Dynamics}

\author{Zhida Song}

\affiliation{Beijing National Laboratory for Condensed Matter Physics and Institute
of Physics, Chinese Academy of Sciences, Beijing 100190, China}

\author{Jimin Zhao}

\affiliation{Beijing National Laboratory for Condensed Matter Physics and Institute
of Physics, Chinese Academy of Sciences, Beijing 100190, China}

\author{Zhong Fang}

\affiliation{Beijing National Laboratory for Condensed Matter Physics and Institute
of Physics, Chinese Academy of Sciences, Beijing 100190, China}

\affiliation{Collaborative Innovation Center of Quantum Matter, Beijing, 100084,
China}

\author{Xi Dai}

\email{daix@aphy.iphy.ac.cn}

\affiliation{Beijing National Laboratory for Condensed Matter Physics and Institute
of Physics, Chinese Academy of Sciences, Beijing 100190, China}

\affiliation{Collaborative Innovation Center of Quantum Matter, Beijing, 100084,
China}
\begin{abstract}

In the present paper, we propose that the chiral magnetic effect, the direct
consequence of the presence of  Weyl points in the band structure, can be detected by its coupling to certain phonon
modes, which behave like pseudo scalars under point group transformations. Such coupling can
generate resonance between intrinsic plasmon oscillation and the corresponding phonon modes,
leading to dramatic modification of the optical response by the external magnetic field,
which provides a new way to study chiral magnetic effect by optical measurements.

\end{abstract}
\maketitle

\paragraph{Introduction.}

Weyl semimetal (WSM) \cite{Weyl_1929, nielsen_adler-bell-jackiw_1983, volovik_universe_2003, murakami_phase_2007, wan_topological_2011, burkov_weyl_2011} is a special
type of metallic phase, which contains Weyl points (WPs), the crossing
points of two non-degenerate bands, in its band structure near the
Fermi level. The uniqueness of WSM lies in two different aspects,
the appearance of unclosed Fermi surfaces for the surface states, Fermi arcs \cite{wan_topological_2011, weng_weyl_2015, huang_weyl_2015, potter_quantum_2014,
lv_observation_2015, lv_observation_spin_2015, xu_discovery_2015, xu_discovery_2015-1};
and the abnormal response under the external fields, which can be
summarised as chiral magnetic effect (CME) \cite{fukushima_chiral_2008,son_berry_2012,zyuzin_topological_2012,chen_axion_2013,vazifeh_electromagnetic_2013}
and anomalous Hall effect \cite{fang_anomalous_2003,xiao_berry_2010,yang_quantum_2011}.
Compared to anomalous Hall effect, the CME only occurs in WSM and
attracts lots of research interests in recent years.

The origin of CME can be understood from the basic electromagnetic (EM)
response of a single WP. Under the external magnetic field, the low
energy states around a single WP carry a finite current along the
field direction due to the chirality of the zeroth landau band \cite{nielsen_adler-bell-jackiw_1983}.
It is clear that the net CME current under a static magnetic field must be zero for any equilibrium
states, which can be guaranteed by the cancellation of the contributions
from different WPs with opposite chiralities, and the CME can only manifest
itself in the non-equilibrium states driven by the external fields
\cite{vazifeh_electromagnetic_2013, chen_axion_2013}. For instance, the external electric
field parallel to the magnetic field will pump the electrons between
WPs with opposite chiralities, which is known as chiral anomaly \cite{nielsen_adler-bell-jackiw_1983,stephanov_chiral_2012,son_berry_2012},
leading to particle imbalance and hence a non-zero net CME current.
In the DC limit, such electron pumping will be eventually balanced by inter-valley
scattering process caused by impurities. And the CME/chiral anomaly
leads to negative magneto-resistance along the magnetic field direction,
which has been studied intensively in recent years \cite{son_chiral_2013, kim_boltzmann_2014, burkov_chiral_2014, burkov_negative_2015, kim_dirac_2013, huang_observation_2015, xiong_evidence_2015, zhang_signatures_2016, li_giant_2015, li_negative_2016}.
In the present Letter, however, we will focus on the CME in the AC
limit, which will cause exotic coupling between some specific phonon
mode and the CME current under the magnetic field. As will be explained
in details in the following, such abnormal coupling will cause a dramatic
change in the optical properties, i.e. the reflectivity, which can be detected by optical
measurement directly.

\paragraph{The pseudo scalar phonon. }

\begin{figure}
\begin{centering}
\includegraphics[width=1\linewidth]{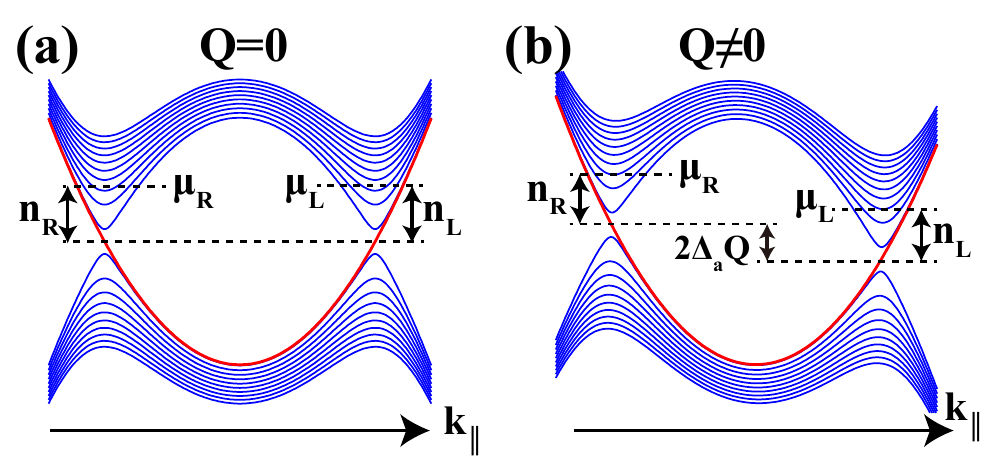}
\par\end{centering}

\protect\caption{\label{fig:Q-CME}Landau bands and deformation potential in a two
WPs model. The red and blue lines represent the zeroth (chiral) Landau
band and other Landau bands respectively. Chiral charge density is
defined as $n_{a}=n_{R}-n_{L}$. (a) is the case of $Q=0$ and (b)
is the case of $Q\protect\neq0$. }
\end{figure}

According to CME, a current proportional to the magnetic field can be generated
by the imbalanced chemical potentials between the WPs with opposite chiralities.
The simplest way to understand CME is to consider a multi-Weyl system
with different chemical potentials for each WPs, where the chiral current
can be expressed as \cite{son_berry_2012}:
\begin{equation}
\mathbf{J}_{\mathrm{CME}}=\frac{e^{2}\mathbf{B}}{4\pi^{2}\hbar^{2}}\sum_{\mathbf{K}_{i}}\chi_{i}\mu_{i}\label{eq:j_CME}
\end{equation}
In the above equation, $\mathbf{K}_{i}$, $\chi_{i}$ and $\mu_{i}$
denote the position, chirality and chemical potential of the $i$-th
WP, respectively.  Eq. (\ref{eq:j_CME}) can be illustrated
through a simplest model containing only two WPs shown in Fig. (\ref{fig:Q-CME}).
Under the external magnetic field, the low energy electronic states
near the WPs become one dimensional Landau bands, which can only disperse
along the field direction as shown in Fig. (\ref{fig:Q-CME}). Since
the imbalance of the chemical potentials only occurs between different
valleys, it is apparent that only the zeroth Landau band contributes
to the the chiral current, which can be written as:

\begin{equation}
\mathbf{J}_{\mathrm{CME}}=-\frac{e^{2}\mathbf{B}}{4\pi^{2}\hbar}\int_{\mathrm{occ}}dk_{\parallel}\frac{\partial\epsilon_{0}\left(k_{\parallel}\right)}{\hbar\partial k_{\parallel}}=\frac{e^{2}\mathbf{B}}{4\pi^{2}\hbar^2}\left(\mu_{R}-\mu_{L}\right)
\end{equation}
where $k_{\parallel}$ is the wavevector along $\mathbf{B}$, $e\left|\mathbf{B}\right|/\left(4\pi^{2}\hbar\right)$
is the degeneracy of each Landau level state and $\int_{\mathrm{occ}}$
integrates over all the occupied states.

It is well known that the imbalanced chemical potentials can be induced
by chiral anomaly effect \cite{nielsen_adler-bell-jackiw_1983,stephanov_chiral_2012,son_berry_2012},
which pumps electrons between WPs with opposite chiralities. In the
present Letter, we propose that, besides chiral anomaly, such imbalanced
chemical potentials and hence the chiral current can also be induced
by the deformation potentials associated with some optical phonon modes.
Since the generic WSM contains multiple pairs of WPs, it is important
to know that for a given phonon mode the contributions from different
pairs of WPs will cancel each other or not, which is fully determined
by the crystal symmetry. In the main text we will mostly focus on
the conclusions of such group theory analysis and leave the mathematical
proof in the supplementary \cite{supp}. For the sake of simplicity, in the current
study we only consider the non-degenerate optical phonon modes in
the long wavelength limit, which form one dimensional representations
of the little group at the $\Gamma$ point. To the leading order,
the electron-phonon coupling for a specific phonon mode can be expressed
as:

\begin{equation}
\hat{H}_{ep} = \frac{1}{V} \sum_{\mathbf{K}_{i}\mathbf{p}} \hat{\psi}_{\mathbf{K}_{i}+\mathbf{p}}^{\dagger} \Delta_{\mathbf{K}_{i},Q} \hat{\psi}_{\mathbf{K}_{i}+\mathbf{p}}Q\label{eq:Hep}
\end{equation}
which describes the energy shifts of the WPs by the deformation potential
owing to the phonon mode $Q$.
The summation over $\mathbf{p}$ only concerns the low energy states around
 each WP.
In the present Letter, we consider
the AC limit and together with the clean limit, where the relaxation rate
due to the impurity scattering is much less than the typical phonon
or plasmon frequency. Therefore, for simplicity we neglect
the impurity scattering and only consider the dynamical effect.
Consider that the system deforms from $Q=0$ (Fig. (\ref{fig:Q-CME}) (a))
to $Q\neq0$ (Fig. (\ref{fig:Q-CME}) (b)). Since the carrier relaxation
is neglected here, the occupation of the Landau bands does not follow
the change of the Hamiltonian, leading to the imbalanced chemical potential
$\mu_{i}=\Delta_{\mathbf{K}_{i},Q}Q$. Then according to Eq. (\ref{eq:j_CME}),
the CME current induced by the deformation potential caused by the
phonon mode $Q$ can be obtained as:

\begin{equation}
\mathbf{J}_{\mathrm{CME},Q}=\frac{N_{W}e^{2}\mathbf{B}}{4\pi^{2}\hbar^{2}}\Delta_{a,Q}Q\label{eq:j_CME_Q}
\end{equation}
where $\Delta_{a,Q}=\frac{1}{N_{W}}\sum_{i}^{N_{W}}\chi_{i}\Delta_{\mathbf{K}_{i},Q}$
and $N_{W}$ is the number of WPs.

A phone mode $Q$ can couple to the CME current only if $\Delta_{a,Q}$
does not vanish, which sets the constrain for the allowed phonon modes.
From the symmetry point of view, all the WSM materials can be divided
into two classes by whether or not their space groups contain improper
rotations, such as the inversion or mirror operators. As proved rigorously
in the supplementary, if the space group does not contain any improper
rotations, the symmetry allowed phonon modes are those $A_{1}$ modes,
which behave as scalars under the little group at $\Gamma$ point.
Up to now most of the WSM materials do have improper rotations in
the space groups and thus belong to the latter class. In this class,
the phonon modes that couple to CME must behave as pseudo scalars
under the point group transformations. In other words, these phonon
modes must keep identity under all the proper rotations and change
signs under all the improper rotations. Such phonon modes are called
pseudo scalar phonons in this Letter, which can be either polarised
or unpolarised. Apparently, if a pseudo scalar mode is
polarised its polarisation vector must be parallel (perpendicular)
to any symmetry axes (mirrors) of the point group. As a result, as long
as there are two or more non-parallel axes or mirror normals $Q$
can not be polarised. The above conclusions can be easily derived
from the invariance of the Hamiltonian (\ref{eq:Hep}) under all the
point group transformations, which is described in details in the supplementary
\cite{supp}. We apply this criterion to some Dirac semimetals and
WSMs studied recently in the literature and find that in $\mathrm{HgCr_{2}Se_{4}}$
\cite{xu_chern_2011}, $\mathrm{Cd_{3}As_{2}}$ \cite{wang_three-dimensional_2013}
and $A\mathrm{Mo_{3}}X_{3}$ \cite{gibson_three-dimensional_2015}
the symmetry allowed phonon modes do exist \cite{supp}.
In the following, we will illustrate how the coupling between the pseudo scalar
phonon and the CME affects the lattice dynamics and hence manifests itself
in the optical response.

\paragraph{Equations of motion (EOMs)}

From Eq. (\ref{eq:j_CME_Q}), we can find that in WSM a current along
the external magnetic field $\mathbf{B}$ can be induced by the motion
of the pseudo scalar phonon mode. Such a CME current will generate
charge accumulation at the sample boundaries and cause an internal electric
field $\mathbf{E}$, which leads to inter-valley electron pumping
under the magnetic field $\mathbf{B}$ due to the chiral anomaly.
Therefore in such a system, the EOMs for pseudo scalar phonon mode
$Q$, chiral charge density defined as $n_{a}=\sum_{i}^{N_{W}}\chi_{i}n_{i}$ (Fig. (\ref{fig:Q-CME}) (b)),
as well as the electric field $\mathbf{E}$ are all coupled
together, leading to quite unusual EM responses in WSM. In the present
Letter, we will mainly focus on the EOMs for the unpolarised phonon
mode and briefly discuss the possible effects related to the polarised
phonon modes thereafter. Tracing out the electronic degrees of freedom
in Eq. (\ref{eq:Hep}), we obtain the effective Lagrangian density
for the phonon as:
\begin{equation}
\mathcal{L} = \frac{M_{ph}}{2\Omega} \dot{Q}^2 - \frac{M_{ph}\omega_{ph}^{2}}{2\Omega} Q^{2} - \Delta_{a}n_{a}Q
\end{equation}
where $M_{ph}$ is the effective mass, $\omega_{ph}$ is the phonon
frequency and $\Omega$ is the unit cell volume. Then the EOM of $Q$
can be obtained as:
\begin{equation}
 \ddot{Q} + \omega_{ph}^{2}Q+\frac{\Delta_{a}\Omega}{M_{ph}}n_{a}=0\label{eq:EOM-Q-np}
\end{equation}
On the other hand, according to Ref. \cite{stephanov_chiral_2012,son_berry_2012}
the low energy electronic dynamics near the WPs leads to the chiral anomaly,
the electron pumping between WPs with opposite chiralities, which can
be expressed as:

\begin{equation}
\dot{n}_{a}=\frac{e^{2}N_{W}}{4\pi^{2}\hbar^{2}}\mathbf{E}\cdot\mathbf{B}\label{eq:EOM-na}
\end{equation}

Next we consider the EOM for the EM wave propagating in the WSM,
which can be written as
\begin{equation}
\ddot{\mathbf{D}} + \dot{\mathbf{J}} + \frac{1}{\mu}\nabla\left(\nabla\cdot\mathbf{E}\right)-\frac{1}{\mu}\nabla^{2}\mathbf{E}=0\label{eq:EOM-EM}
\end{equation}
In WSM $\mathbf{D}$ can be simply set as $\epsilon_{0}\kappa\mathbf{E}$
where $\kappa$ is dielectric constant in the high frequency limit.
While the current $\mathbf{J}$ contains multiple terms with different
origins, which can be written as $\mathbf{J}_{\mathrm{total}} = \mathbf{J}_{\mathrm{op}} + \mathbf{J}_{\mathrm{H}} + \mathbf{J}_{\mathrm{CME}} + \mathbf{J}_{\mathrm{AH}}$.
The first two terms are the optical current $\mathbf{J}_{\mathrm{op},i}=\sigma_{i}\left(\omega,\mathbf{q}\right)\mathbf{E}_{i}$
(where $i$ distinguishes the transversal and longitudinal responses) and
the Hall current $\mathbf{J}_{\mathrm{H}}=\sigma_{\mathrm{H}}\left(\omega,\mathbf{q}\right)\mathbf{E}\times\mathbf{B}$,
which exist in all metallic systems. In addition, there are two
novel components in a WSM, the CME and anomalous Hall currents. In the clean
limit, the imbalanced chiral charges (\ref{eq:EOM-na}) have little
chance to be scattered from the zeroth Landau bands to the other Landau bands,
because the scattering rate is much less than the corresponding frequency. Thus
the imbalance of the ``chemical potentials'', or the highest occupied
energies, of the zeroth Landau bands can be written as $\Delta\mu_{0}=N_{W}\Delta_{a}Q+n_{a}/\nu_{D}$,
where $\nu_{D}=\frac{e\left|\mathbf{B}\right|}{4\pi^{2}\hbar^{2}v_{F}}$
is the density of states of a single zeroth Landau band. The two terms contained
in $\Delta\mu_{0}$ describe the pseudo scalar phonon contribution and
chiral anomaly contribution, respectively. Thus the total CME current
is explicitly written as:
\begin{equation}
\mathbf{J}_{\mathrm{CME}}=\frac{e^{2}\mathbf{B}}{4\pi^{2}\hbar^{2}}\left(N_{W}\Delta_{a}Q+\frac{n_{a}}{\nu_{D}}\right)
\end{equation}
The anomalous Hall current can be expressed as $\mathbf{J}_{\mathrm{AH}}=\frac{e^{2}}{\hbar}\mathbf{E}\times\mathbf{d}$,
where $\mathbf{d}=\int\frac{d^{3}\mathbf{k}}{\left(2\pi\right)^{3}} n_F(\epsilon_\mathbf{k})\boldsymbol{\Omega}_{\mathbf{k}}$ is the integral of the Berry curvature contributed by all the occupied bands over the Brillouin zone.
Including all the four different origins of the current in Eq. (\ref{eq:EOM-EM}),
we obtain the coupled equations (\ref{eq:EOM-Q-np}), (\ref{eq:EOM-na})
and (\ref{eq:EOM-EM}) for the pseudo scalar phonon mode $Q$, chiral
charge density $n_{a}$, and the electric field $\mathbf{E}$, which
can be solved under different conditions leading to several novel
optical properties introduced below.

\begin{figure}
\begin{centering}
\includegraphics[width=1\linewidth]{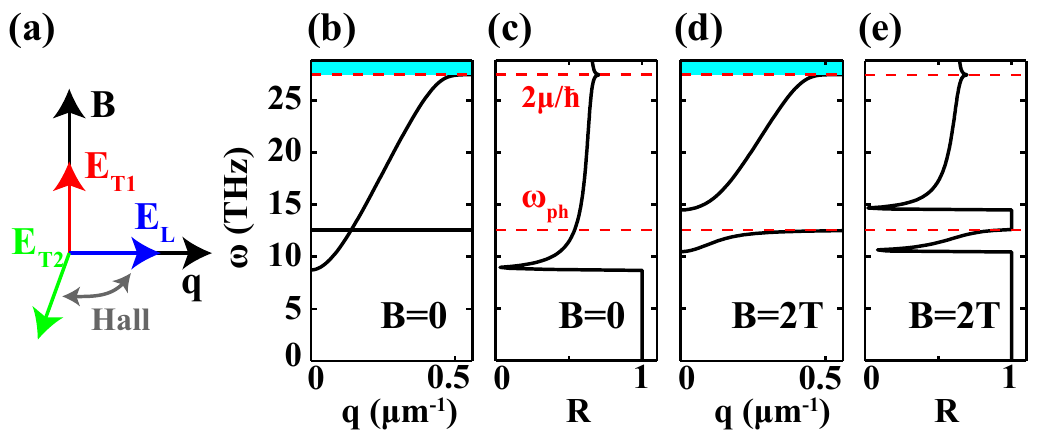}
\par\end{centering}

\protect\caption{\label{fig:EM-trans}EM wave coupled with unpolarised pseudo scalar
phonon. (a) illustrates the fields configuration, which ensures that
the mode coupled with CME current ($E_{T1}$) and is free from Hall
effects. (b) and (c) are the numerically calculated dispersion and
reflectivity for $E_{T1}$ at $\left|\mathbf{B}\right|=0$. (d) and
(e) are the dispersion and reflectivity at $\left|\mathbf{B}\right|=2\mathrm{T}$.
The cyan areas are the inter-band single particle excitation zones \cite{ashby_chiral_2014}.}
\end{figure}

\begin{figure}
\begin{centering}
\includegraphics[width=1\linewidth]{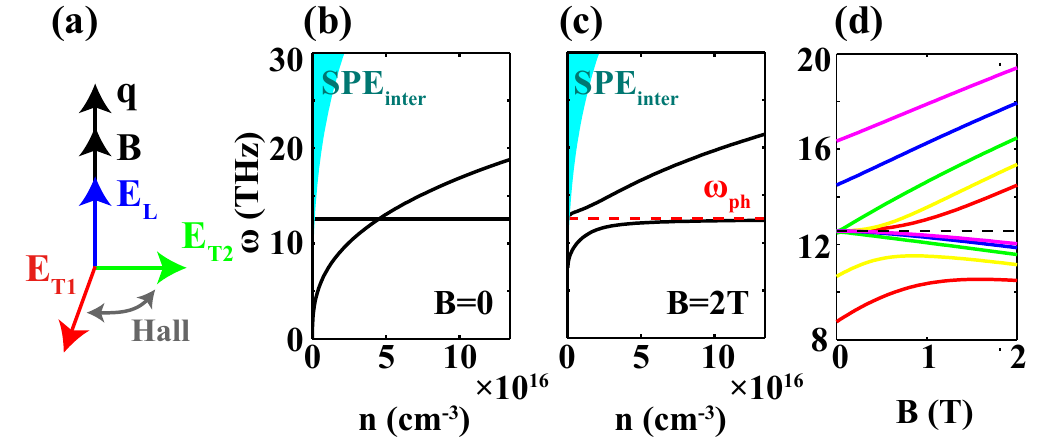}
\par\end{centering}

\protect\caption{\label{fig:EM-long}Plasmon coupled with unpolarised pseudo scalar
phonon. (a) illustrates the fields configuration, which ensures that
the mode coupled with CME current ($E_{L}$) is free from Hall effect.
(b) and (c) are the eigen frequencies of the coupled modes as functions of carrier density at
$\left|\mathbf{B}\right|=0$ and $\left|\mathbf{B}\right|=2\mathrm{T}$, respectively,
where the cyan areas are the inter-band single particle excitation
zones \cite{lv_dielectric_2013}. In (d) we plot the eigen frequencies
of the coupled modes as functions of the magnetic field.
Different colored lines represent different intrinsic plasmon frequencies
(carriers densities).}
\end{figure}

\paragraph{Physical consequences.}
The above EOMs give a new mechanism to couple the EM wave with the pseudo scalar phonons through CME, which
only exists under the external magnetic field.
If the particular pseudo scalar phonon mode is polarised, the additional coupling caused by CME only modifies the coupling constant between
the two. While if the  pseudo scalar phonon mode is unpolarised, it can now couple to the EM wave by the CME, which will generate a dramatic change
in the optical response near the phonon frequency. In other words, the originally optical inactive phonon mode can become active under the external
magnetic field due to the CME. In this present Letter, we will mainly focus on the unpolarised phonon modes and only briefly discuss the polarised
phonon modes thereafter, because the effect in the former case is more dramatic. To avoid additional contributions from both ordinary and
anomalous  Hall effects, we always set the external magnetic field $\mathbf{B}$ to be parallel to the vector $\mathrm{\mathbf{d}}$
(if there is a nonzero $\mathbf{d}$) and consider the polarisation of the EM mode to be along the magnetic field. In total there are two EM modes,
which are free from Hall effects and will be studied in detail below.

The first one is the transverse mode describing an EM wave with wavevector $\mathbf{q}$ being perpendicular  and
polarisation direction being parallel to the magnetic field $\mathbf{B}$, as shown in Fig. (\ref{fig:EM-trans}) (a).
The EOM for the electric field in such situation can be simplified as
\begin{align}
 & -\kappa\omega^{2}E_{T1}-i\frac{\omega}{\epsilon_{0}}\sigma_{T}\left(\omega\mathbf{q}\right)E_{T1}+\frac{1}{\mu\epsilon_{0}}\mathbf{q}^{2}E_{T1}\nonumber \\
 & -i\frac{\omega e^{2}B}{4\pi^{2}\epsilon_{0}\hbar^{2}}\left(N_{W}\Delta_{a}Q+\frac{n_{a}}{\nu_{D}}\right)=0\label{eq:EOM-EM-trans}
\end{align}
where $\sigma_{T}$ denotes the transverse optical conductivity \cite{ashby_chiral_2014}.
The effective dielectric function in this case can be obtained by
substituting the solution of Eq. (\ref{eq:EOM-Q-np}) and (\ref{eq:EOM-na})
into Eq. (\ref{eq:EOM-EM-trans}), which can be written as{\small{}
\begin{align}
\epsilon_{r,T} & =\kappa+i\frac{\sigma_{T}\left(\omega\mathbf{q}\right)}{\omega\epsilon_{0}}-\frac{e^{4}B^{2}}{16\pi^{4}\hbar^{4}\epsilon_{0}\omega^{2}}\left[\frac{N_{W}}{\nu_{D}}+\frac{N_{W}^{2}\Delta_{a}^{2}\Omega}{M_{ph}\left(\omega^{2}-\omega_{ph}^{2}\right)}\right]\nonumber \\
 & =\epsilon_{r,T}^{0}-\frac{e^{4}B^{2}}{16\pi^{4}\hbar^{4}\epsilon_{0}\omega^{2}}\left[\frac{N_{W}}{\nu_{D}}+\frac{N_{W}^{2}\Delta_{a}^{2}\Omega}{M_{ph}\left(\omega^{2}-\omega_{ph}^{2}\right)}\right]\label{eq:eps_r_trans}
\end{align}
}where $\epsilon_{r,T}^{0}$ is the original dielectric function without
magnetic field. The above solution indicates that through CME the
photon with the polarisation along the field direction can hybridize
with the unpolarised pseudo scalar phonon to form a special type of
polariton mode, which can only exist under the magnetic field. If the phonon frequency $\omega_{ph}$ is
higher than the plasmon frequency $\omega_{pl}$, as shown in Fig. (\ref{fig:EM-trans}) (b),
without the magnetic field there is a level crossing between $E_{T1}$ and $Q$
at frequency $\omega_{ph}$ and momentum $\hbar\mathbf{q}_{0}$, which becomes anti-crossing under the
magnetic field through the CME. By
solving the equation $\epsilon_{r,T}=c^{2}\mathbf{q}^{2}/\omega^{2}$
to the first order of $\mathbf{B}$ and $\Delta_{a}$, we can get
the hybridization gap between the photon like and phonon like branches
to be
\begin{align}
\Delta\omega & \approx\frac{e^{2}N_{W}\Delta_{a}\left|\mbox{\ensuremath{\mathbf{B}}}\right|}{\pi^{2}\hbar^{2}}\nonumber \\
 & \times\sqrt{\frac{\Omega}{8\epsilon_{0}M_{ph}\omega_{ph}^{3}\left(\frac{2\epsilon_{r,T}^{0}}{\omega_{ph}}+\frac{\partial\epsilon_{r,T}^{0}}{\partial\omega}\right)_{\omega_{ph},\mathbf{q}_{0}}}}
\end{align}
The existence of such CME induced polariton mode can be inferred from
the dramatic modification of the reflectivity under the magnetic field,
which can be written as $R=\left|\frac{\sqrt{\epsilon_{r}}-1}{\sqrt{\epsilon_{r}}+1}\right|^{2}$
\cite{ashcroft_solid_1976}. For example, as $\omega$ approaches
$\omega_{ph}$ the last term in Eq. (\ref{eq:eps_r_trans}) will diverge,
leading to an additional peak in the reflectivity.

We further apply numerical calculations for the CME induced polariton modes
using the typical parameters
for the semiconductor or semimetal materials, which are $N_{W}=4$, $v_{F}=2.2\times10^{5}\mathrm{m/s}$,
$\kappa=15$, $\Delta_{a}=8\mathrm{eV/\AA}$, $M_{ph}=90\mathrm{u}$,
$\omega_{ph}=12.6\mathrm{THz}$, and $\Omega=2021\mathrm{\AA}^{3}$. The Fermi momentum  has been chosen as
$p_{F}=0.0063\mathrm{\AA}^{-1}$.
The dispersion of the polariton modes and the reflectivity of light
are calculated and shown in Fig. (\ref{fig:EM-trans}). A CME induced hybridization gap between the phonon and photon-like
branches can be found clearly under the magnetic field , as shown in Fig. (\ref{fig:EM-trans}) (d), which generates an additional
peak in the optical reflectivity after the plasmon edge, as illustrated in Fig. (\ref{fig:EM-trans}) (e).

The second EM mode, that is free from the Hall effects, is the longitudinal or plasmonic mode \cite{panfilov_density_2014} with wavevector and polarisation direction both
being parallel to the magnetic field.
The EOM  can then be simplified to
\begin{align}
 & -\kappa\omega^{2}E_{L}-i\frac{\omega}{\epsilon_{0}}\sigma_{L}\left(\omega\mathbf{q}\right)E_{L}\nonumber \\
 & -i\frac{\omega e^{2}B}{4\pi^{2}\epsilon_{0}\hbar^{2}}\left(N_{W}\Delta_{a}Q+\frac{n_{a}}{\nu_{D}}\right)=0\label{eq:EOM-EM-long}
\end{align}
where $\sigma_{L}$ denotes the longitudinal conductivity. The effective
dielectric function for $E_{L}$ is similar with Eq. (\ref{eq:eps_r_trans})
except that $\sigma_{T}$ is replaced by $\sigma_{L}$.
Similar to the polariton case, the CME will generate coupling between
the unpolarised pseudo scalar phonon and the plasmon modes, which
are otherwise completely decoupled without the CME. Similarly we can obtain
the hybridization gap between the two branches as
\begin{equation}
\Delta\omega\approx\frac{e^{2}N_{W}\Delta_{a}\left|\mbox{\ensuremath{\mathbf{B}}}\right|}{\pi^{2}\hbar^{2}}\sqrt{\frac{\Omega}{8\epsilon_{0}M_{ph}\omega_{ph}^{3}\kappa\left.\frac{\partial\epsilon_{r,L}^{0}}{\partial\omega}\right|_{\omega_{ph},\mathbf{q}_{0}}}}\label{eq:Det_omega_pl}
\end{equation}
where $\epsilon_{r,L}^{0}$ is the original longitudinal dielectric function without
magnetic field \cite{lv_dielectric_2013}. The similar numerical calculations for the coupled plasmon and phonon modes
have been carried out using the same parameters introduced above and the results are shown in Fig. (\ref{fig:EM-long}),
where the hybridization between the plasmon and phonon modes can be induced only with the finite magnetic field.
Such mixed modes under magnetic field can be detected by the Raman scattering with the proper setup.


\begin{figure}
\begin{centering}
\includegraphics[width=1\linewidth]{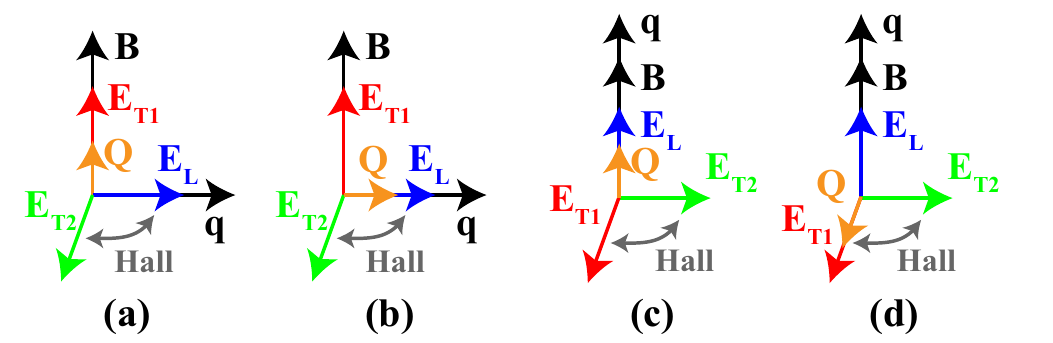}
\par\end{centering}

\protect\caption{\label{fig:EM-pol}EM modes coupled with polarised pseudo scalar
phonon.}
\end{figure}

In the end, we briefly discuss the situation for the polarised pseudo
scalar phonon modes. We list the four different configurations in
Fig. (\ref{fig:EM-pol}) as examples and always set as $\mathbf{B}\parallel\mathbf{d}$
if there is a nonzero $\mathbf{d}$. In Fig. (\ref{fig:EM-pol}) (a),
the phonon mode couples to the EM mode $E_{T1}$ because their polarisation
directions are parallel to each other. The CME  enhances such
coupling and modifies the dispersion of the polariton modes in this
case. The situation in Fig. (\ref{fig:EM-pol}) (b) is more complicated.
The two transverse EM modes, the longitudinal plasmon mode and the
pseudo scalar phonon mode are all coupled together under the magnetic
field due to the combination of the Hall effect and CME. The case
in Fig. (\ref{fig:EM-pol}) (c) is similar with the case in Fig. (\ref{fig:EM-pol})
(a), in which the CME enhances the coupling strength between $Q$ and
$E_{L}$. The case in Fig. (\ref{fig:EM-pol}) (d) is as complicated
as the case in Fig. (\ref{fig:EM-pol}) (b) because all the modes
are coupled together.

\paragraph{Conclusions. }

In summary, a special type of phonon mode, which behaves as pseudo scalar
under improper rotation, is proposed to be coupled to the EM dynamics in the WSMs or Dirac
semimetals. If such a pseudo scalar phonon mode is unpolarised, it will become optically
active under the external magnetic field, manifesting the CME in these materials. Such CME
induced phonon-EM wave coupling can be detected easily by the optical reflectivity spectroscopy
near the phonon frequency, providing a new experimental way to study the CME in topological
semimetals.

\paragraph{Acknowledgments. }
We acknowledge the supports from National Natural Science Foundation of China,
the National 973 program of China (Grant No. 2013CB921700) and the ``Strategic Priority Research Program (B)"
of the Chinese Academy of Sciences (Grant No. XDB07020100).

\bibliographystyle{apsrev4-1}
\bibliography{Weyl_Ph}

\newpage{}~
\newpage{}

\clearpage


\section*{Supplementary material}

# Supplemental Material for "Detecting Chiral Magnetic Effect by Lattice Dynamics"

Zhida Song,[1] Jimin Zhao,[1] Zhong Fang,[1,2] and Xi Dai[1,2,*]

[1]*Beijing National Laboratory for Condensed Matter Physics and Institute of Physics, Chinese Academy of Sciences, Beijing 100190, China*
[2]*Collaborative Innovation Center of Quantum Matter, Beijing, 100084, China*

## I. PSEUDO SCALAR PHONON AND CME CURRENT

In this section we will prove that only pseudo scalar phonon can couple with the CME current. Assume the interaction between a single band electronic system and a non-degenerate phonon mode $Q$ in long wavelength limit is:

$$\hat{H}_{ep} = \frac{1}{V} \sum_{\mathbf{K}_i, \mathbf{p}} \hat{\psi}^\dagger_{\mathbf{K}_i+\mathbf{p}} \Delta_{\mathbf{K}_i,Q} \hat{\psi}_{\mathbf{K}_i+\mathbf{p}} Q \tag{S1}$$

where $\mathbf{K}_i$ and $\Delta_{\mathbf{K}_i,Q}$ denote the position and deformation potential of the $i$-th WP, respectively, and the summation over $\mathbf{p}$ only concerns the low energy states around each WP. According to the main text, the CME current caused by the deformation potential due to the motion of phonon is:

$$\mathbf{J}_{\mathrm{CME},Q} = \frac{N_W e^2 \mathbf{B}}{4\pi^2 \hbar^2} \Delta_{a,Q} Q \tag{S2}$$

where $N_W$ is the number of WPs and $\Delta_{a,Q}$ is defined as $\Delta_{a,Q} = \frac{1}{N_W} \sum_{\mathbf{K}_i} \chi_i \Delta_{\mathbf{K}_i,Q}$, in which $\chi_i$ is the chirality of the $i$-th WP. The necessary and sufficient condition for $Q$ to couple with CME current is $\Delta_{a,Q} \neq 0$.

On one hand, $Q$ must form a one dimensional representation of the little group at $\Gamma$ (denoted as $G$):

$$\forall g \in G, \quad gQg^{-1} = D_Q(g) Q \tag{S3}$$

On the other hand, the invariance of $\hat{H}_{ep}$ under group $G$ implies:

$$\frac{1}{V} \sum_{\mathbf{K}_i, \mathbf{p}} \hat{\psi}^\dagger_{\mathbf{K}_i+\mathbf{p}} \Delta_{\mathbf{K}_i,Q} \hat{\psi}_{\mathbf{K}_i+\mathbf{p}} Q$$
$$= \frac{1}{V} \sum_{\mathbf{K}_i, \mathbf{p}} g \hat{\psi}^\dagger_{\mathbf{K}_i+\mathbf{p}} \Delta_{\mathbf{K}_i,Q} \hat{\psi}_{\mathbf{K}_i+\mathbf{p}} Q g^{-1}$$
$$= \frac{1}{V} \sum_{\mathbf{K}_i, \mathbf{p}} \hat{\psi}^\dagger_{g\mathbf{K}_i+\mathbf{p}} \Delta_{\mathbf{K}_i,Q} \hat{\psi}_{g\mathbf{K}_i+\mathbf{p}} gQg^{-1} \tag{S4}$$

Combining Eq. (S3) and Eq. (S4) we have:

$$\Delta_{g\mathbf{K}_i,Q} = D_Q(g) \Delta_{\mathbf{K}_i,Q} \tag{S5}$$

Assume there is a right hand WP at $\mathbf{K}$, by symmetry principle we know there will be a right (left) hand WP at $g\mathbf{K}$ if $\det g = 1$ ($\det g = -1$). Thus the quantity $\Delta_{a,Q}$ is simplified to

$$\Delta_{a,Q} = \frac{1}{N_W} {\sum_{\mathbf{K}_i}}' \chi_i \frac{1}{|H^{\mathbf{K}_i}|} \sum_{g \in G} \det g \cdot \Delta_{g\mathbf{K}_i,Q}$$
$$= \frac{1}{N_W} {\sum_{\mathbf{K}_i}}' \chi_i \Delta_{\mathbf{K}_i,Q} \frac{1}{|H^{\mathbf{K}_i}|} \sum_{g \in G} \det g \cdot D_Q(g) \tag{S6}$$

where the summation $\sum'$ goes through all non-equivalent WPs and $H^{\mathbf{K}_i}$ is the little group of $\mathbf{K}_i$. The denominator $|H^{\mathbf{K}_i}|$ in Eq. (S6), which is the order of $H^{\mathbf{K}_i}$, is used to cancel the multiplicity from $\mathbf{K}_i$-invariant operations. Since $\det g$ itself is an irreducible representation of $G$, orthogonality theorem ensures that only pseudo scalar phonons, i.e. $D_Q(g) = \det g$, have non-vanishing $\Delta_{a,Q}$.

It should be noticed that, for a group only consists of proper rotations, $\det g = 1$ is just the identity representation and "pseudo scalars" reduce to scalars.

## II. PSEUDO SCALAR PHONONS IN MATERIALS

In this section we will discuss the pseudo scalar phonons in materials.

Firstly we give a criterion for the existence of pseudo scalar phonons. As phonon modes are completely determined by crystal structure, we can give a criterion based only on symmetry principles. Assume there are $L$ non-equivalent occupied Wyckoff sites, each of which has a site symmetry group (SSG) $H^a$ ($a = 1 \cdots L$). For a given $H^a$, $G$ can be decomposed as its cosets:

$$G = s_1 H^a + s_2 H^a + \cdots \tag{S7}$$

For the $a$-th kind of Wyckoff sites, $H^a$ keeps each site invariant and $s_i$ interchanges them, so there are $\frac{|G|}{|H^a|}$ atoms in this kind. Thus the mechanical representation formed by the displacements configuration is:

$$D(g) = \bigoplus_{a=1}^{L} \left[ D^a_{\mathrm{int}}(s) \otimes D_{O(3)}(h) \right] \tag{S8}$$

where:

$$s \cdot h = g \qquad s \in G/H^a \qquad h \in H^a \tag{S9}$$

and $D^a_{\mathrm{int}}(s)$ is the $\frac{|G|}{|H^a|} \times \frac{|G|}{|H^a|}$ interchange matrix of the $a$-th kind atoms, $D_{O(3)}(h)$ is just the O(3) rotation matrix

| | Materials | Space Group | Little Group at $\Gamma$ | Relevant Wyckoff Sites | SSGs | Pseudo Scalar Phonon [a] | Polarised |
|---|---|---|---|---|---|---|---|
| Weyl — non magnetic | $A\text{Bi}_{1-x}\text{Se}_x\text{Te}_3$ [S1] | 160 | $C_{3v}$ | - | - | - | - |
| | BiTeI under pressure [S1] | 156 | $C_{3v}$ | - | - | - | - |
| | Se/Te under pressure [S2] | 152/153 | $D_3$ | $3a$ | $C_2$ | $A_1$ | No |
| | TaAs [S3, S4] | 109 | $C_{4v}$ | - | - | - | - |
| Weyl — magnetic | $A_2\text{Ir}_2\text{O}_7$ [S5] | 227.131 [b] | $\mathcal{O}_h$ [c] | - | - | - | - |
| | $\text{HgCr}_2\text{Se}_4$ [S6] | 141.557 [b] | $\mathcal{D}_{4h}$ [c] | $8c, 16h$ | $C_{2h}, C_s$ | $2 \times A_{1u}$ | No |
| Dirac — Class I [d] | $\text{Cu}_3\text{PdN}$ [S7] | 221 | $O_h$ | - | - | - | - |
| | $A_3\text{Bi}$ [S8] | 194 | $D_{6h}$ | - | - | - | - |
| | BaAuBi-family [S9] | 194 | $D_{6h}$ | - | - | - | - |
| | LiGaGe-family [S9] | 186 | $C_{6v}$ | - | - | - | - |
| | $\text{SrSn}_2\text{As}_2$ [S9] | 160 | $C_{3v}$ | - | - | - | - |
| | $\text{Cd}_3\text{As}_2$ [S10] | 137 | $D_{4h}$ | $3 \times 8g, 8f$ | $C_s, C_2$ | $4 \times A_{1u}$ | No |
| | $\text{Cd}_3\text{As}_2$ [S10] | 110 | $C_{4v}$ | $9 \times 16b, 2 \times 8a$ | $C_1, C_2$ | $29 \times A_2$ | No |
| Dirac — Class II | $\beta$-cristobalite $\text{BiO}_2$ [S11] | 227 | $O_h$ | - | - | - | - |
| | $\text{HfI}_3$ [S9] | 193 | $D_{6h}$ | - | - | - | - |
| | $A\text{Mo}_3X_3$-family [S9] | 176 | $C_{6h}$ | $2 \times 6h, 2c$ | $C_s, C_{3h}$ | $3 \times A_u$ | Yes |
| | Distorted Spinel [S12] | 74 | $D_{2h}$ | $4a, 4d, 8h, 8i$ | $C_{2h}, C_{2h}, C_s, C_s$ | $4 \times A_u$ | No |

[a] The pseudo scalar representation may have different symbols in different groups. In the first class point groups (consisting of proper rotations), it is just the identity representation $A_1$. While in the second (centrosymmetric) and third (non-centrosymmetric but with improper rotations) class point groups, it is usually referred as $A_{1u}$ and $A_2$, respectively. The prefactor is the number of pseudo scalar phonon modes. "-" means there is no pseudo scalar phonon.

[b] The magnetic space group is referred to Ref. [S13].

[c] The magnetic little group is defined as $\mathcal{O}_h = T_h + \mathcal{T} \cdot (O_h - T_h)$ and $\mathcal{D}_{4h} = C_{4h} + \mathcal{T} \cdot (D_{4h} - C_{4h})$, respectively, where $\mathcal{T}$ is the time reversal operator.

[d] The classification follows Ref. [S14].

TABLE S1: Pseudo scalar phonons in Weyl and Dirac semimetals.

of $h$. Then we can apply orthogonality theorem to calculate the multiplicity of pseudo scalar representation. By recognizing that $D_{\text{int}}^a(s)$ is off-diagonal unless $s$ equals to identity, the multiplicity $\lambda$ can be simplified to:

$$\lambda = \sum_{a=1}^{L} \frac{1}{|G|} \sum_{s \in G/H^a} \sum_{h \in H^a} \det(s \cdot h) \operatorname{tr}\left[D_{\text{int}}^a(s) \otimes D_{\text{O}(3)}(h)\right]$$

$$= \sum_{a=1}^{L} \frac{1}{|H^a|} \sum_{h \in H^a} \det(h) \cdot \operatorname{tr}\left[D_{\text{O}(3)}(h)\right] \quad \text{(S10)}$$

which indicates that $\lambda$ is determined only by the SSGs of occupied Wyckoff sites. The SSGs can contribute to Eq. (S10) are easy to exhaust and we classify them into two classes:

$$\mathcal{H}_3 = \{C_1, C_i\} \quad \text{(S11)}$$

$$\mathcal{H}_1 = \{C_2, C_3, C_4, C_6, C_{2h}, C_{3h}, C_{4h}, C_{6h}, C_s, S_4, S_6\} \quad \text{(S12)}$$

The symbol for point group follows Ref. [S15]. Each group in $\mathcal{H}_3$ will contribute 3 in Eq. (S10) and each group in $\mathcal{H}_1$ will contribute 1 in Eq. (S10). This can be understood in a quite intuitive perspective: as a pseudo scalar mode the displacement of each atom must be parallel (perpendicular) to any symmetry axes (mirrors) in the SSG, so as long as there are two or more non-parallel axes or mirror normals the displacement must be zero. (Symmetry axis here represent both rotation and rotation-reflection axis). Indeed $\mathcal{H}_3$ are those groups have no symmetry axes or mirrors and $\mathcal{H}_1$ are those groups have only one direction of symmetry axes or mirror normals. Therefore we achieve the criterion: the existence of pseudo scalar phonons is equivalent with the existence of occupied Wyckoff sites with SSGs in $\mathcal{H}_3 \cup \mathcal{H}_1$. And the number of pseudo scalar modes can be simply counted as the number of non-equivalent Wyckoff sites with SSGs in $\mathcal{H}_1$ plus three times of the number of non-equivalent Wyckoff sites with SSGs in $\mathcal{H}_3$.

We list the pseudo scalar phonons in some recently studied WSMs and Dirac semimetals in table (S1).



\end{document}